\documentclass[journal,twoside,web]{ieeecolor}
\usepackage{jabbrv}
\usepackage{generic}
\usepackage{cite}
\usepackage{amsmath,amssymb,amsfonts}
\usepackage{algorithmic}
\usepackage{graphicx}
\usepackage{algorithm,algorithmic}
\usepackage{textcomp}
\usepackage{multirow}
\usepackage{booktabs} 
\usepackage{makecell}
\usepackage{xr}
\usepackage{xcolor}
\usepackage{array}
\usepackage{placeins}

\usepackage{generic}
\usepackage{cite}
\usepackage{amsmath,amssymb,amsfonts}
\usepackage{algorithmic}
\usepackage{graphicx}
\usepackage{textcomp}
\usepackage{multirow}
\usepackage{booktabs}
\def\BibTeX{{\rm B\kern-.05em{\sc i\kern-.025em b}\kern-.08em
    T\kern-.1667em\lower.7ex\hbox{E}\kern-.125emX}}
\begin{document}
\title{Reconstructing Unseen Sentences from Speech-related Biosignals for Open-vocabulary Neural Communication}
\author{Deok-Seon Kim, Seo-Hyun Lee, Kang Yin, and Seong-Whan Lee, \IEEEmembership{Fellow, IEEE}
\thanks{This work was supported in part by the Institute of Information and Communications Technology Planning and Evaluation (IITP) grant funded by the Korea Government [Ministry of Science and ICT (MSIT)] under Grant RS-2019-II190079, in part by the Artificial Intelligence Graduate School Program (Korea University) under Grant RS-2021-II21206, in part by the Artificial Intelligence Innovation Hub and AI Technology for Interactive Communication of Language Impaired Individuals under Grant RS-2024-00336673, and in part by the National Research Foundation of Korea (NRF) funded by the MSIT (MetaSkin: Developing Next-generation Neurohaptic Interface Technology that Enables Communication and Control in Metaverse by Skin Touch) under Grant 2022-2-00975. \emph{(Corresponding author: Seong-Whan Lee.)}}
\thanks{Deok-Seon Kim, Kang Yin, and Seong-Whan Lee are with the Department of Artificial Intelligence, Korea University, Seoul 02841, South Korea (e-mail: deokseon\_kim@korea.ac.kr; charles\_kang@korea.ac.kr; sw.lee@korea.ac.kr). Seo-Hyun Lee is with the Department of Brain and Cognitive Engineering, Korea University, Seoul 02841, South Korea (e-mail: seohyunlee@korea.ac.kr).}}

\maketitle
 \begin{abstract}

Brain-to-speech (BTS) systems represent a groundbreaking approach to human communication by enabling the direct transformation of neural activity into linguistic expressions. While recent non-invasive BTS studies have largely focused on decoding predefined words or sentences, achieving open-vocabulary neural communication comparable to natural human interaction requires decoding unconstrained speech. Additionally, effectively integrating diverse signals derived from speech is crucial for developing personalized and adaptive neural communication and rehabilitation solutions for patients. This study investigates the potential of speech synthesis for previously unseen sentences across various speech modes by leveraging phoneme-level information extracted from high-density electroencephalography (EEG) signals, both independently and in conjunction with electromyography (EMG) signals. Furthermore, we examine the properties affecting phoneme decoding accuracy during sentence reconstruction and offer neurophysiological insights to further enhance EEG decoding for more effective neural communication solutions. Our findings underscore the feasibility of biosignal-based sentence-level speech synthesis for reconstructing unseen sentences, highlighting a significant step toward developing open-vocabulary neural communication systems adapted to diverse patient needs and conditions. Additionally, this study provides meaningful insights into the development of communication and rehabilitation solutions utilizing EEG-based decoding technologies.

\end{abstract}


\section{Introduction}
\label{sec:introduction}

\IEEEPARstart{B}{rain-computer} interface (BCI) bridges the gap between human brain activity and computer devices, enabling users to communicate or control systems directly through decoded intentions\cite{chaudhary2016brain, gao2021interface}. 
BCI systems have shown potential in multiple domains, including the diagnosis of brain disorders~\cite{prabhakar2020framework, bagherzadeh2025automated}, communication systems~\cite{milekovic2018stable, sereshkeh2017eeg}, functional rehabilitation for individuals with movement disorders~\cite{cho2021neurograsp, samejima2021brain}, sleep-stage classification~\cite{yu2022mrasleepnet, bao2024feature}, and emotion decoding~\cite{kim2015abstract}. Particularly, brain-to-speech (BTS) stands out as a groundbreaking method for brain signal-driven communication that translates neural signals into speech~\cite{anumanchipalli2019speech, lee2020neural, angrick2021real, defossez2023decoding}. 

Speech is a fundamental component of human interaction, therefore, BTS systems aim to provide intuitive communication solutions for individuals with impaired speech capabilities. Human speech generates various signals depending on the mode of production, including acoustic signals and biosignals such as electroencephalography (EEG) and electromyography (EMG). Overt speech, for instance, yields various signals---acoustic, EEG, and EMG---while whispered speech incorporates EEG and EMG signals along with less distinctive acoustic signals. In contrast, imagined speech yields speech-related EEG signals while excluding movement-related signals such as EMG or even acoustic signals. These distinctions make certain signals more suitable for specific clinical applications. For example, overt and whispered speech can be practical for individuals with partial speech abilities, whereas imagined speech can be crucial for patients with severe paralysis or conditions such as locked-in syndrome, which prevents any form of physical articulation.

Recent studies have shown great potential in reconstructing speech by decoding brain signals, mostly collected in invasive ways during overt or attempted speech~\cite{makin2020machine, moses2021neuroprosthesis, liu2023decoding, wairagkar2025instantaneous}. However, efforts have increasingly focused on decoding brain signals from whispered and imagined speech~\cite{saha2019speak, lee2021decoding} in non-invasive methods for the cases of a broader range of individuals. Considering the significance of BTS systems for patients, the adaptation of non-invasive methods is crucial due to their safety, cost-effectiveness, and accessibility compared to invasive techniques, which involve surgical procedures that can pose significant risks~\cite{metzger2023high, willett2023high, duraivel2023high}. Optimally applying different speech modes to the model is expected to enhance the effectiveness and accessibility of the communication system while addressing diverse clinical requirements and the unique conditions of individual patients. 
 
Among non-invasive modalities, EEG has emerged as a prominent tool for decoding speech-related features due to its high temporal resolution. However, most EEG-based studies have been limited to classifying or reconstructing words within predefined categories, which restricts the system’s flexibility and applicability in real-world communication~\cite{kim2023diff, cooney2020evaluation, dekker2023dais, alonso2025pronounced}. To enable more practical and meaningful communication, sentence-level reconstruction is necessary, as word decoding often lacks the semantic richness required for effective interaction. This requires generating unconstrained sentences from speech-related signals, supporting an open-vocabulary approach rather than relying on predefined classes. Previous studies have attempted to decode unseen words or sentences, demonstrating the feasibility of generalizing beyond fixed vocabularies. For instance,~\cite{lee2023AAAI} explored voice reconstruction for unseen words, showing that the model could generate intelligible speech signals beyond the training set vocabulary. Similarly, EEG-to-text decoding studies~\cite{wang2022open, amrani2024deep} ensured that test sentences were entirely novel, evaluating the model’s ability to handle open-vocabulary scenarios. However, achieving this level of complexity with EEG alone is particularly challenging due to its low signal-to-noise ratio (SNR) and high variability. Therefore, flexibly configuring the use of multimodal data to each patient condition could enable BTS systems to provide more adaptive and reliable means of communication.

In this study, we demonstrate the feasibility of reconstructing unseen sentences through sentence-level speech synthesis by leveraging phoneme-level information from non-invasive biosignals of different conditions. In addition, we analyze factors influencing sentence reconstruction performance and perform neurophysiological evaluations to investigate insights for further improving EEG decoding. This study highlights the potential of an unconstrained communication system that accommodates diverse patient needs and conditions by incorporating multiple modalities, such as EEG and EMG, and exploring various speech modes, including overt, whispered, and imagined speech.

\section{Dataset Description}

\subsection{Data Acquisition} 
Fifteen healthy subjects (10 males and 5 females; mean age: 26.73 $\pm$ 4.41 years; 8 Korean-native speakers and 7 English-native speakers) were recruited for the experiment. The study was carried out under the Declaration of Helsinki and approved by the Korea University Institutional Review Board [KUIRB-2022-0079-03]. All participants signed an informed consent form before the experiment. EEG and EMG data were recorded with Brain Vision/Recorder (BrainProduct GmbH, Germany) at a sampling rate of 1,000~Hz. EEG data were collected with the Standard 128-channel actiCAP (reference: FCz), which follows the 10\%-System with additional electrodes positioned based on the 5\%-System to achieve higher spatial resolution. EMG data were recorded from 10 channels placed on the face and neck, with the detailed placement illustrated in Supplementary Fig.~S1. Vocal sounds were simultaneously recorded using a microphone with a sampling rate of 44,100 Hz, ensuring synchronization with the biosignals.

\begin{figure}[t!]
\centerline{\includegraphics[width=\columnwidth]{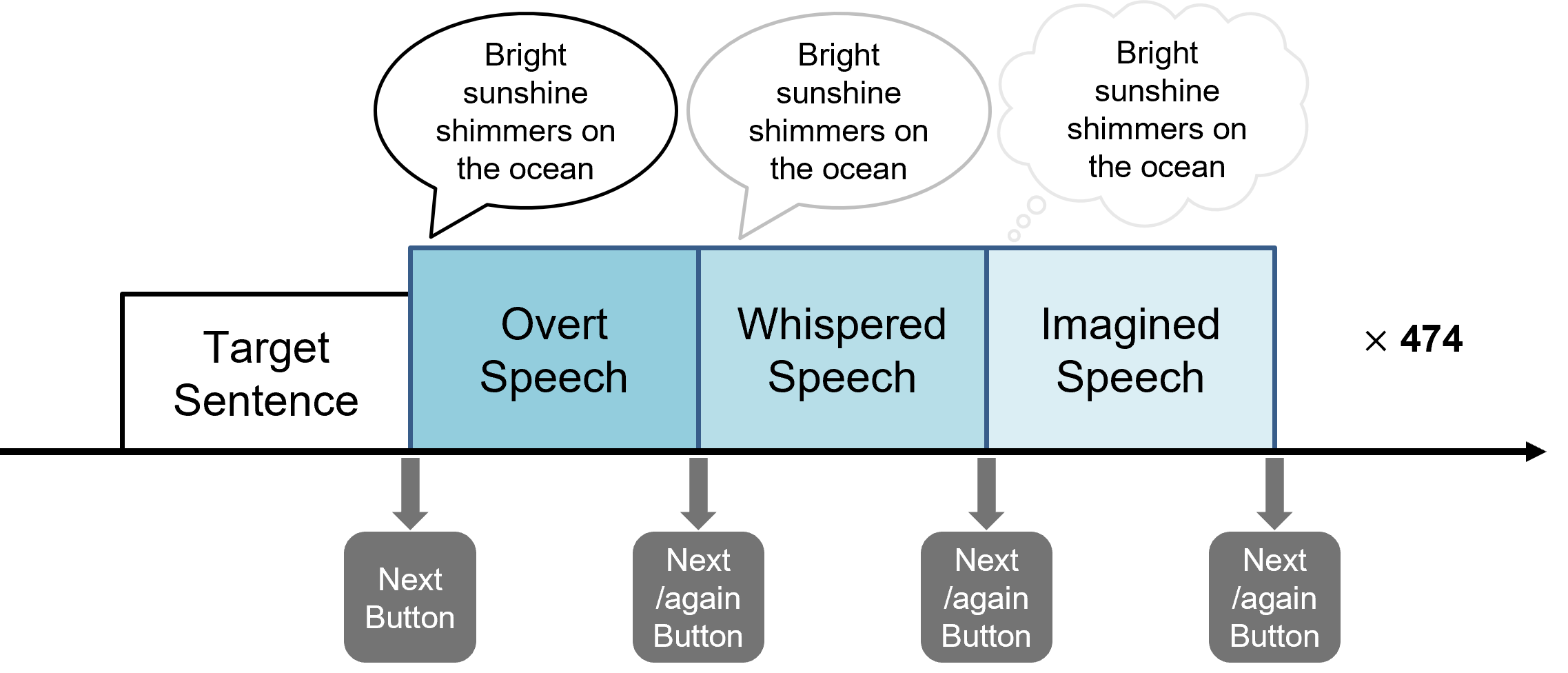}}
\caption{Experimental paradigm with three different speech modes: overt, whispered, and imagined speech. EEG, EMG, and audio signals of 474 different sentences including various phonemes were recorded.}
\label{experiment}
\end{figure}

As illustrated in Fig.~\ref{experiment}, each subject performed three distinct speech modes in sequence for each of the 474 sentences: overt speech, where they naturally vocalized the sentence aloud; whispered speech, involving the quiet mouthing of the sentence without full vocalization; and imagined speech, where they silently thought of speaking without producing any vocalization. All sentences were in English and comprised 460 sentences from the MOCHA-TIMIT corpus~\cite{wrench2000multichannel}, known to be designed for phonetically balanced data, along with 14 additional sentences. These sentences were transcribed using a total of 39 phoneme classes and one silence class. A summary of sentence-level phoneme statistics across subjects can be found in Supplementary Table~S1.

Our study is designed with potential applications for patients, where participation in prolonged experiments may not be feasible. To accommodate these constraints, we collected data using only a single trial per sentence, unlike many prior studies that collect multiple trials for the same word or sentence. In each trial, a speech target sentence was displayed on a screen as text (e.g., ``Dolphins are intelligent marine mammals'') as in Table~\ref{tab1}. Participants had the option to press a `next' button to proceed to the subsequent speech paradigm or press an `again' button to repeat the current trial. Although all trials were recorded, our preprocessing included an epoching step in which only the final instance after the last ‘again’ press was retained for the experiment and analysis. A total of 474 trials of each speech mode were collected and split into training and test datasets, comprising 444 and 30 trials respectively, with no overlap or repetition between the datasets. The sentences used in our test set were unseen sentences that were not present in the training dataset.

\begin{table}[t!]
\caption{Example of Target Sentences and Phonetic Transcriptions}
\centering
\renewcommand{\arraystretch}{1.2}
\begin{tabular}{p{0.96\linewidth}}
\toprule
\textbf{Example of Target Sentences and Transcriptions} \\ \toprule
\textbf{Did you eat lunch yesterday?} \\
{/sil, d, ih, d, y, uw, iy, t, l, ah, n, ch, y, eh, s, t, er, d, ey, sil/} \\ \midrule
\textbf{Few people live to be a hundred} \\
{/sil, f, y, uw, p, iy, p, ah, l, l, ih, v, t, ah, b, iy, ah, hh, ah, n, d, r, ih, d, sil/} \\ \midrule
\textbf{How permanent are their records?} \\
{/sil, hh, aw, p, er, m, ah, n, ah, n, t, aa, r, dh, eh, r, r, ah, k, ao, r, d, z, sil/} \\ \midrule
\textbf{The emperor had a mean temper} \\
{/sil, dh, iy, eh, m, p, er, er, hh, ae, d, ah, m, iy, n, t, eh, m, p, er, sil/} \\ \bottomrule
\end{tabular}
\label{tab1}
\end{table}

\begin{figure*}[t!]
\centerline{\includegraphics[width=\textwidth]{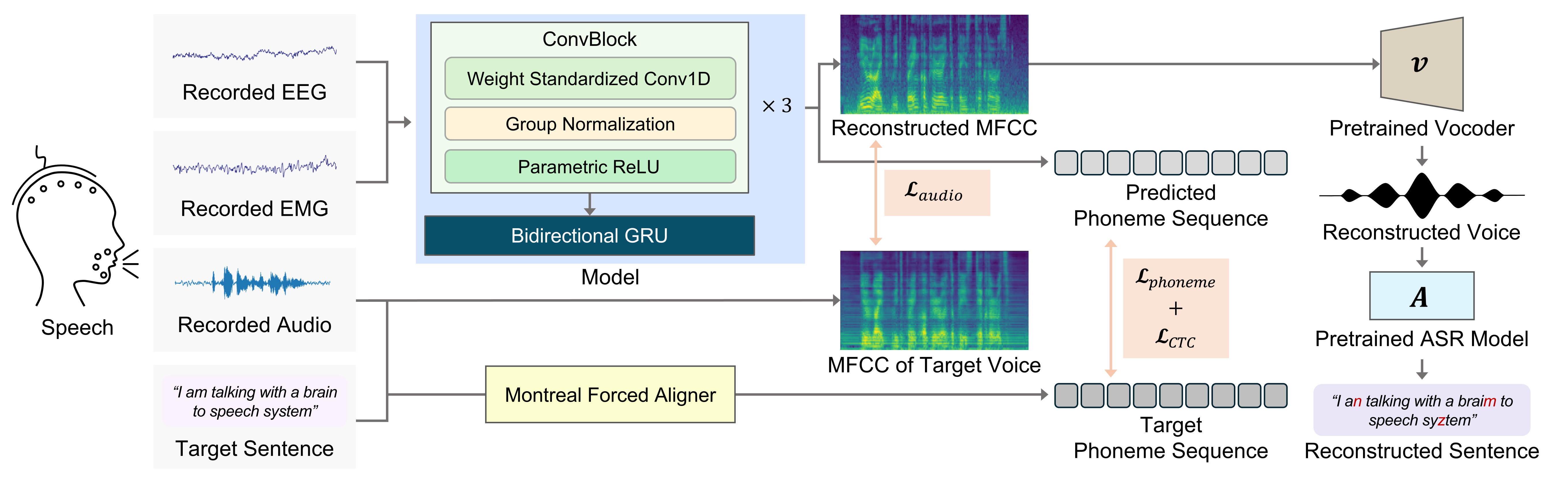}}
\caption{The overall decoding framework. EEG data, or EEG combined with EMG data, is provided as the input, obtained from different modes of speech (overt, whispered, and imagined speech). The model comprises three Convblocks followed by Bi-GRU. The model generates the reconstructed MFCC and phoneme sequences. A pre-trained vocoder ($v$) synthesizes audio from the predicted MFCC and a pre-trained ASR model ($A$) converts the reconstructed voice into text.}
\label{framework}
\end{figure*}
\subsection{Data Preprocessing} 
Several preprocessing steps were applied to EEG and EMG data using EEGLAB~\cite{delorme2004eeglab} and MNE-python~\cite{esch2018mne}. For EEG data, a band-pass filter between 0.5--200~Hz was applied, while for EMG data, a high-pass filter with a cutoff of 2~Hz was used. Additionally, a notch filter was utilized for both EEG and EMG signals to remove 60~Hz power line noise and its harmonics, which are typically introduced by electrical devices and can significantly contaminate the signals. After these filtering steps, common average referencing was applied separately to the EEG and EMG data to reduce noise and improve signal quality. EEG and EMG signals were epoched into segments corresponding to the actual duration of the speech tasks, with baseline correction by subtracting the average value of the 200~ms before each trial.

\section{Methodology}

\subsection{Framework} 

The overall framework of our study is shown in Fig.~\ref{framework}. The overt speech recordings were used as ground truth (GT) for mel-frequency cepstral coefficients (MFCC) generation. Sentence texts were transformed into phoneme sequences and temporally aligned with the audio recordings using the Montreal forced aligner (MFA), ensuring that the MFCCs and phoneme sequences were synchronized. The phoneme sequences were represented using a total of 40 classes, consisting of 39 distinct phonemes and one silence class, and served as GT for phoneme sequence prediction. Given that whispered speech has an inferior vocal signal and imagined speech lacks a reference voice, overt speech samples recorded using the same sentences were employed as target audio~\cite{lee2023AAAI, gaddy2020digital}. Biosignals recorded across three speech modes---overt, whispered, and imagined---were used as input to the model. Depending on the setup, they consisted of either EEG alone or a combination of EEG and EMG, where EEG and EMG signals were concatenated along the channel dimension, enabling the model to leverage the complementary information from both modalities. To account for individual differences in biosignal characteristics, we developed a personalized decoding model that is specifically trained and optimized for each subject. The model consists of three convolutional blocks (ConvBlocks) and a bidirectional gated recurrent unit (Bi-GRU). Each ConvBlock incorporates weight standardization, group normalization, and parametric rectified linear unit (PReLU) activation to improve training stability and enhance feature extraction. The output of the ConvBlocks is fed into the Bi-GRU, which captures both forward and backward temporal dependencies. The model produces two outputs: sentence-level MFCC and the corresponding phoneme sequence. Each output is associated with a specific training loss.

During inference, the trained model predicts sentence-level MFCC from input biosignals. The recorded audio and the target sentence were not given to the model during the inference phase. To transform predicted audio features into audible speech, we utilize the pre-trained HiFi-GAN~\cite{kong2020hifi}, which generates high-fidelity audio from feature representations. The generated speech was converted into text using the DeepSpeech~\cite{hannun2014deep} model, a pre-trained automatic speech recognition (ASR) model.

\subsection{Training Loss Terms} 
During training, we adopted two different training loss strategies depending on the speech modes: $\mathcal{L}_{\text{overt}}$ and $\mathcal{L}_{\text{silent}}$.
\subsubsection{Overt Speech}

 For overt speech, the training loss $\mathcal{L}_{\text{overt}}$ is composed of three components: $\mathcal {L}_{\text{audio}}$, $\mathcal {L}_{\text{phoneme}}$ and $\mathcal{L}_{\text{CTC}}$, and is defined as: 
\begin{equation}
{\mathcal { L }}_{\text{overt}}= \alpha\cdot{\mathcal { L }}_{\text{audio}}+{\mathcal { L }}_{\text{phoneme}}+{\mathcal { L }}_{\text{CTC}}.
\end{equation}

We set $\alpha$ to 0.5 to scale the loss value and ensure a balanced contribution of each loss term, preventing any single loss from dominating the training process due to magnitude differences.

The audio loss, $\mathcal {L}_{\text{audio}}$, is computed as the Euclidean distance between the recorded and predicted MFCC audio features:
\begin{equation}
\mathcal{L}_{\text {audio}}=\sum_{i=1}^{N}\left\|y_i-\hat{y}_i\right\|_2,
\end{equation}
where ${y}_i$ represents the $i$-th element of the GT of MFCC feature vector, and $\hat{y}_i$ denotes the corresponding predicted MFCC feature vector, and $N$ is the number of feature dimensions.

The phoneme loss, $\mathcal{L}_{\text{phoneme}}$, is calculated using the cross-entropy as follows:
\begin{equation}
{\mathcal { L }}_{\text{phoneme}} = -\sum_{c=1}^{M} p_{c} \log(q_{c}),
\end{equation}
where $p_{c}$ is the true probability of the class $c$, and $q_{c}$ is the model-predicted probability of the phoneme class, and $M$ is the total number of phoneme classes.

The connectionist temporal classification (CTC) loss, $\mathcal{L}_{\text{CTC}}$, widely employed in ASR systems~\cite{graves2006connectionist}, was used to measure the discrepancy between the predicted and actual phoneme sequence without requiring explicit alignment information.

\subsubsection{Whispered and Imagined Speech}
Since we utilized the corresponding overt speech audio as a reference for the same sentence, alignment between the audio features and the predicted audio features is necessary for whispered and imagined speech. The training loss combines the dynamic time warping (DTW) loss~\cite{gaddy2021improved} and CTC loss, and is defined as:
\begin{equation}
{\mathcal { L }}_{\text{silent}}=\alpha\cdot{\mathcal { L }}_{\text{DTW}}+{\mathcal { L }}_{\text{CTC}}. 
\end{equation}

DTW is used to determine the optimal alignment path between the predicted MFCC and GT. First, a distance matrix, $\mathbf{D}_{\text{dist}}$, is computed using the Euclidean distances between the predicted MFCC and GT. To improve the alignment with the accuracy of the phoneme prediction, we refine the distance matrix by incorporating the log probabilities of phonemes. Specifically, a weighted negative log probability of phoneme predictions, $\log(\mathbf{\hat{P}}^{(i,j)})$, is added to the distance matrix. The resulting cost matrix is formulated as follows:

\begin{equation}
\mathbf{C}_{\text{cost}}^{(i,j)} = \mathbf{D}_{\text{dist}}^{(i,j)} + \beta\cdot (-\log(\mathbf{\hat{P}}^{(i,j)})),
\end{equation}
where $\beta$ is set to 0.5. $i$ and $j$ are index of the matrix. The DTW loss, $\mathcal{L}_{\text{DTW}}$, is then computed as the sum of the costs along this path:
\begin{equation}
\mathcal{L}_{\text{DTW}} = \sum_{i=1}^{T} \mathbf{C}_{\text{cost}}^{(i, {align}(i))},
\end{equation}
where $T$ is the sequence length, and $align(i)$ refers to the optimal alignment at time step $i$.

\subsection{Training Details}
The model was trained using the AdamW optimizer~\cite{loshchilov2019decoupled}, with a weight decay regularization term of $10^{-5}$. The initial learning rate was set to $10^{-3}$ and dynamically adjusted using a cosine annealing learning rate scheduler. To prevent overfitting, we applied early stopping, limited the maximum number of epochs to 200, and incorporated a dropout rate of 0.1 in the Bi-GRU layers.

To handle batching with sequences of varying lengths, we concatenate the biosignals within a batch along the time axis~\cite{gaddy2021improved}. The signals are then reshaped into fixed-length sequences. For a fixed sequence length $l$, given $N_S$ total samples across the batch and $C$ signal channels, the inputs are reshaped to a size of $\left(\left\lceil N_S / l \right\rceil, l, C\right)$ after zero-padding to ensure the total length is a multiple of $l$. Once the network processes the data to generate predicted audio features and phoneme sequences, this operation is reversed to restore the variable-length sequences for alignment and loss computation. In our experiments, a sequence length $l$ of 2,048 is used.

The input to the model consists of either EEG alone or a combination of EEG and EMG. For combined inputs, the total number of input channels is 137, with 127 from EEG and 10 from EMG, resulting in a shape of $(B, C, l)$, where $B$ is the batch size. The data is processed by three ConvBlocks, each performing one-dimensional convolution over the time dimension. The output of the ConvBlocks has a shape of $(B, 512, l/8)$. This is then processed by a bidirectional GRU with a hidden size of 512, resulting in an output shape of $(B, l/8, 1024)$. Finally, this output is passed through two separate linear layers: one produces MFCC features with a shape of $(B, l/8, 80)$, and the other generates phoneme sequence predictions with a shape of $(B, l/8, 40)$.

\subsection{Quantitative and Qualitative Evaluation}
The evaluation was conducted on the outputs from the test set, which consisted of 30 unique sentences that were not used in the training. The root mean square error (RMSE) and mel-cepstral distortion (MCD) were calculated to evaluate the reconstruction quality of the MFCC generated by the model. RMSE was computed after aligning the reconstructed and original MFCC using DTW. To determine the intelligibility of the generated speech, we evaluated the character error rate (CER) of the outputs processed through the ASR model. To assess the performance of the model in accurately reconstructing phoneme sequence, we measured phoneme accuracy, which was further visualized using a confusion matrix. Additionally, we computed the F1-score to compare the ground truth and predicted phoneme sequences, evaluating the model’s effectiveness in phoneme prediction. Furthermore, we used t-distributed stochastic neighbor embedding (t-SNE)~\cite{van2008visualizing} to visualize the average of each phoneme class in the predicted phoneme sequences, illustrating the clustering patterns among phoneme groups.

\subsection{Sentence Properties and Phoneme Decoding Performance Analysis}
To explore the factors influencing phoneme decoding accuracy during sentence reconstruction, we analyzed the relationship between phoneme accuracy and various sentence properties, including phoneme sequence length and W score. 
The W score is defined as the proportion of each phoneme's occurrence in the training dataset relative to the total number of phonemes. Since the silence (`sil') had the highest occurrence count in the dataset, we calculated the W score in two ways: one including `sil' and another excluding `sil' from the total phoneme count. The detailed W scores for all phonemes are provided in Supplementary Table~S2. We visualized the relationship between phoneme accuracy and each sentence property using scatter plots for overt and whispered EEG data. For each scatter plot, we calculated the Pearson correlation coefficient (PCC) to quantify the strength and direction of the linear relationship between phoneme accuracy and the respective sentence property.

\subsection{Neurophysiological Analysis}
To investigate the neurophysiological basis of phoneme decoding performance, we analyzed EEG signals across six standard frequency bands: delta (0.5--4~Hz), theta (4--8~Hz), alpha (8--12~Hz), beta (12--30~Hz), gamma (30--70~Hz), and high gamma (70--200~Hz)~\cite{strijkstra2003subjective, nicolas2012brain, van2008increased, darvas2010high}. For each frequency band, EEG signals were preprocessed using band-pass filtering with a zero-phase FIR filter implemented in MNE to isolate the relevant frequency components. Decoding performance was evaluated for overt, whispered, and imagined speech modes using phoneme accuracy, RMSE, and MCD.

Additionally, source localization was performed using sLORETA~\cite{pascual2002standardized} on EEG signals in the range of 0.5--200~Hz to identify the spatial distribution of brain activity associated with each speech mode. Neural activation patterns were analyzed at 1~s, 1.25~s, 1.5~s after speech onset. The resulting source activation maps were visualized for the dorsal, lateral, and parietal regions, highlighting differences in neural engagement across overt, whispered, and imagined speech modes.

\section{Results}
\subsection{Reconstruction of Unseen Sentences}

\begin{figure}[t!]
\centerline{\includegraphics[width=0.95\columnwidth]{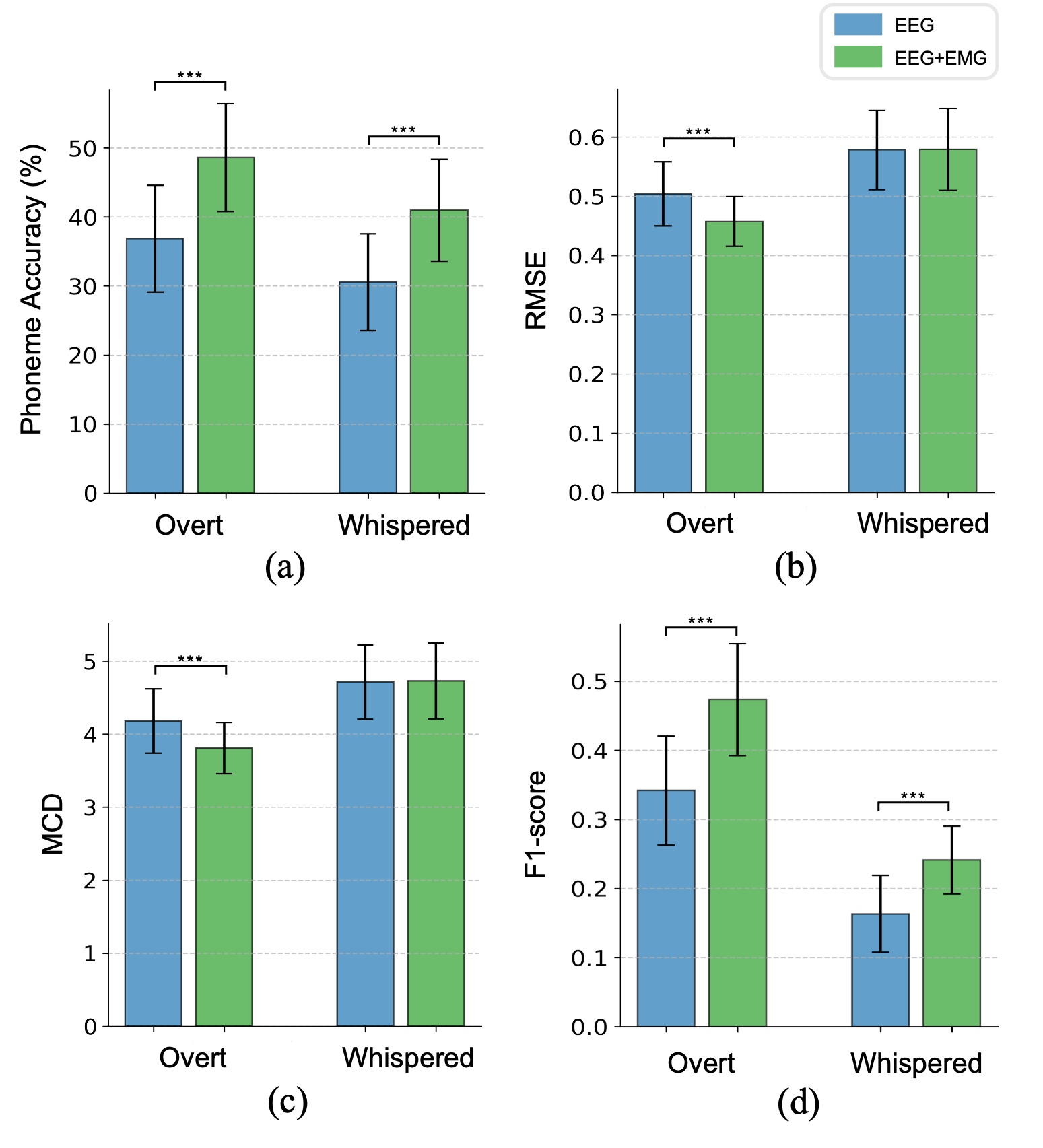}}
\caption{Results of unseen sentence reconstruction. (a) Phoneme accuracy of phoneme sequence, (b) RMSE between the target and reconstructed MFCC, (c) MCD between the original and generated MFCC, (d) F1-score between GT and predicted phoneme sequence. The results are averaged across all 15 participants. The Wilcoxon signed-rank test, a non-parametric statistical method, was performed to evaluate the significance differences in paired data. Significance at the level of $p <$ 0.0001 is indicated by ***.}
\label{performance}
\end{figure}

To evaluate the performance for the generation of unseen sentences, we utilized the test set comprised of unseen sentences not included in the training set. The quantitative evaluation results are displayed in Fig.~\ref{performance}. In overt speech, when combining EEG and EMG data, the phoneme accuracy was 48.58\%, RMSE was 0.46, MCD was 3.81, and the F1-score was 0.47. In whispered speech, the phoneme accuracy reached 40.97\%, with an RMSE of 0.58, MCD of 4.73, and an F1-score of 0.24. When using EEG alone, the phoneme accuracy for overt speech was 36.83\%, RMSE was 0.5, MCD was 4.18, and the F1-score was 0.34. In whispered speech, the phoneme accuracy was 30.54\%, RMSE was 0.58, MCD was 4.71, and the F1-score was 0.16. Detailed subject-wise performance results are presented in Supplementary Table~S3. As phoneme occurrences are not uniformly distributed, we estimated chance performance based on the most frequent phoneme in our dataset. The silence phoneme ``sil" accounted for 11\%, and the decoding results were above this baseline even when phoneme frequency was considered.
Statistical analysis was conducted using the Wilcoxon signed-rank test and significance was evaluated at $p<$ 0.0001. The sample size ($N$=15) corresponds to the number of participants, where each data point represents the average performance across 30 test sentences for each participant. Statistical analysis confirmed that the improvements in all evaluation metrics, including phoneme accuracy, RMSE, MCD, and F1-score were significant for overt speech, while whispered speech exhibited significant improvement in both phoneme accuracy and F1-score.

\begin{table}[t!]
\caption{Examples of Reconstructed Sentences Generated \\by ASR model from Synthesized Audio Derived \\from Biosignals of a Representative Subject}
\centering
\renewcommand{\arraystretch}{1.2}
\begin{tabular}{lllc}
\toprule
 & \textbf{Modality} & \textbf{Sentence}                                               & \textbf{CER} \\
\midrule
\#1 & GT      & \textbf{What a wonderful world}                                             & -            \\
             & Overt    & \textbf{What wonderful w}i\textbf{ld}                                       & 0.18         \\
             & Whispered & One \textbf{wonderful w}inte\textbf{r}                                      & 0.50          \\ 
\midrule
\#2 & GT        & \textbf{I want to be tall as this tree}                                     & -            \\
            &  Overt     & \textbf{I} o\textbf{n}ce me\textbf{t} a \textbf{as this tree}               & 0.36         \\
             & Whispered & \textbf{I want to} r\textbf{e}ad \textbf{tal}e\textbf{s} d\textbf{istr}ict  & 0.40          \\ 
\bottomrule
\end{tabular}
\label{tab2}
\end{table}

\begin{figure*}[t!]
\centerline{\includegraphics[width=0.91\textwidth]{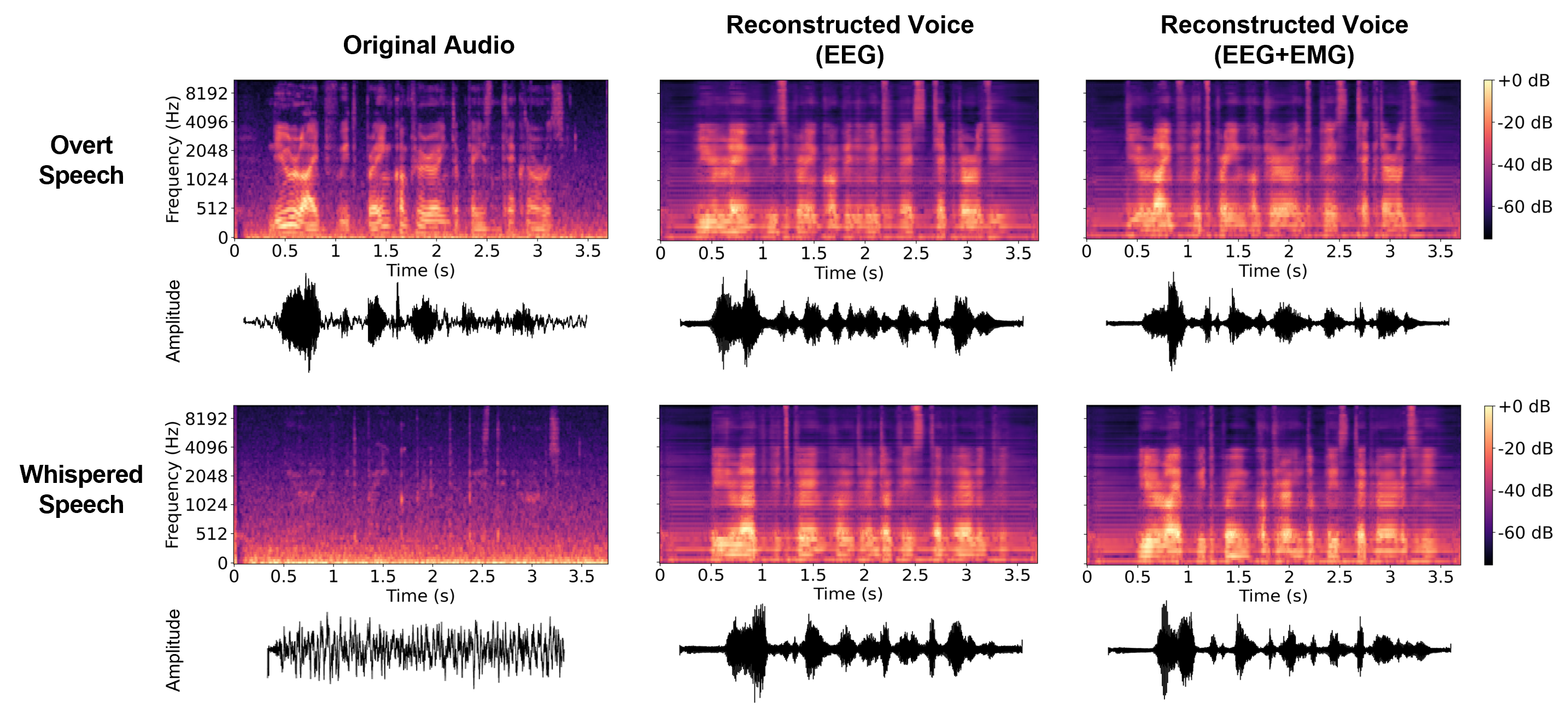}}
\caption{Mel-spectrograms and the audio waveforms of the original voice, the reconstructed voice from biosignals of overt and whispered speech, from a representative subject. The target sentence is ``New life with brain-computer interface technology".}
\label{mel}
\end{figure*}

\begin{figure*}[t!]
\centerline{\includegraphics[width=0.91\textwidth]{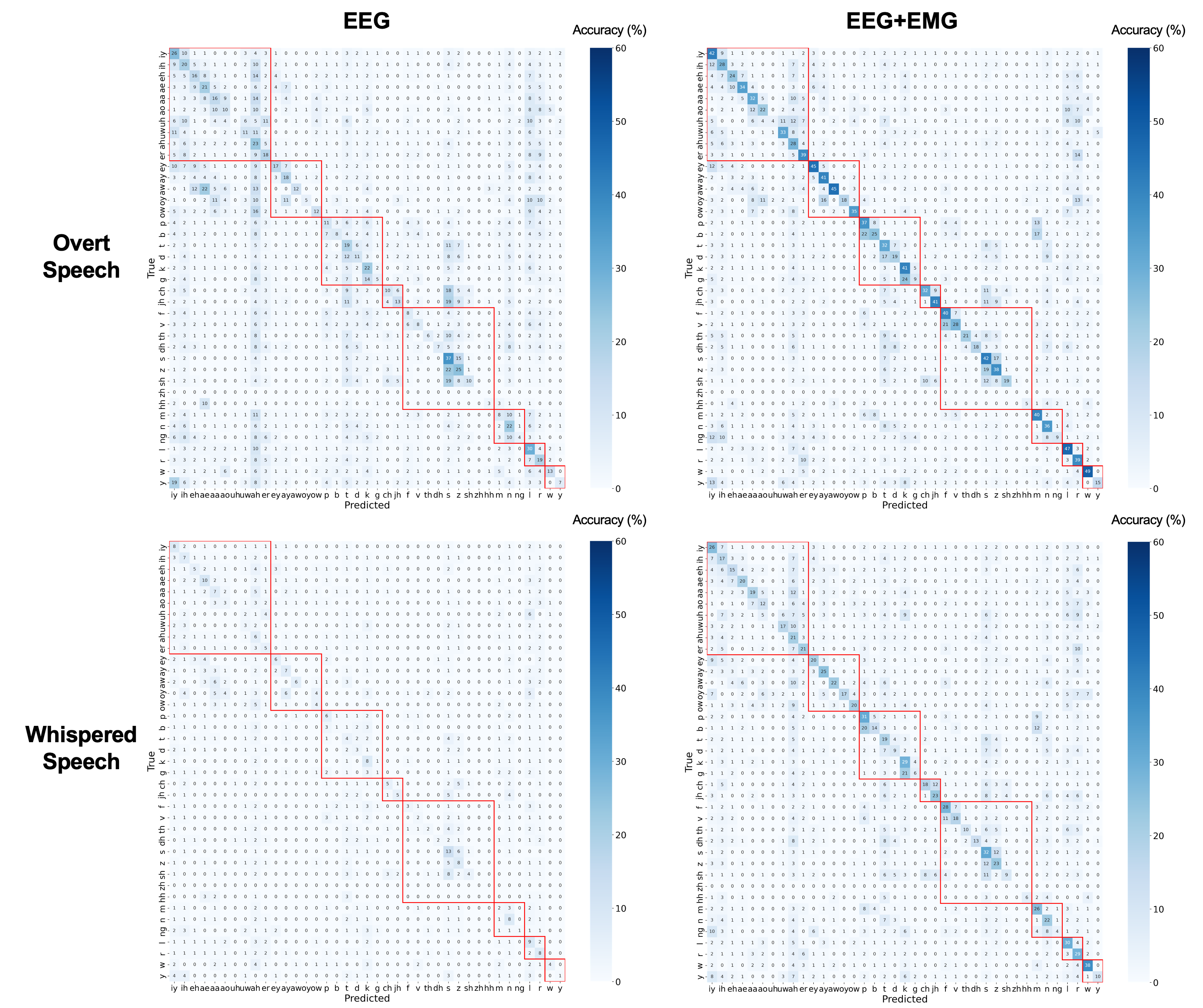}}
\caption{Confusion matrix of predicted phoneme sequence aggregated across all participants. The matrix compares the predicted phonemes with GT for both overt and whispered speech modes. Red boxes highlight predefined phoneme groups.}
\label{confusion}
\end{figure*}

\begin{figure}[t!]
\centerline{\includegraphics[width=\columnwidth]{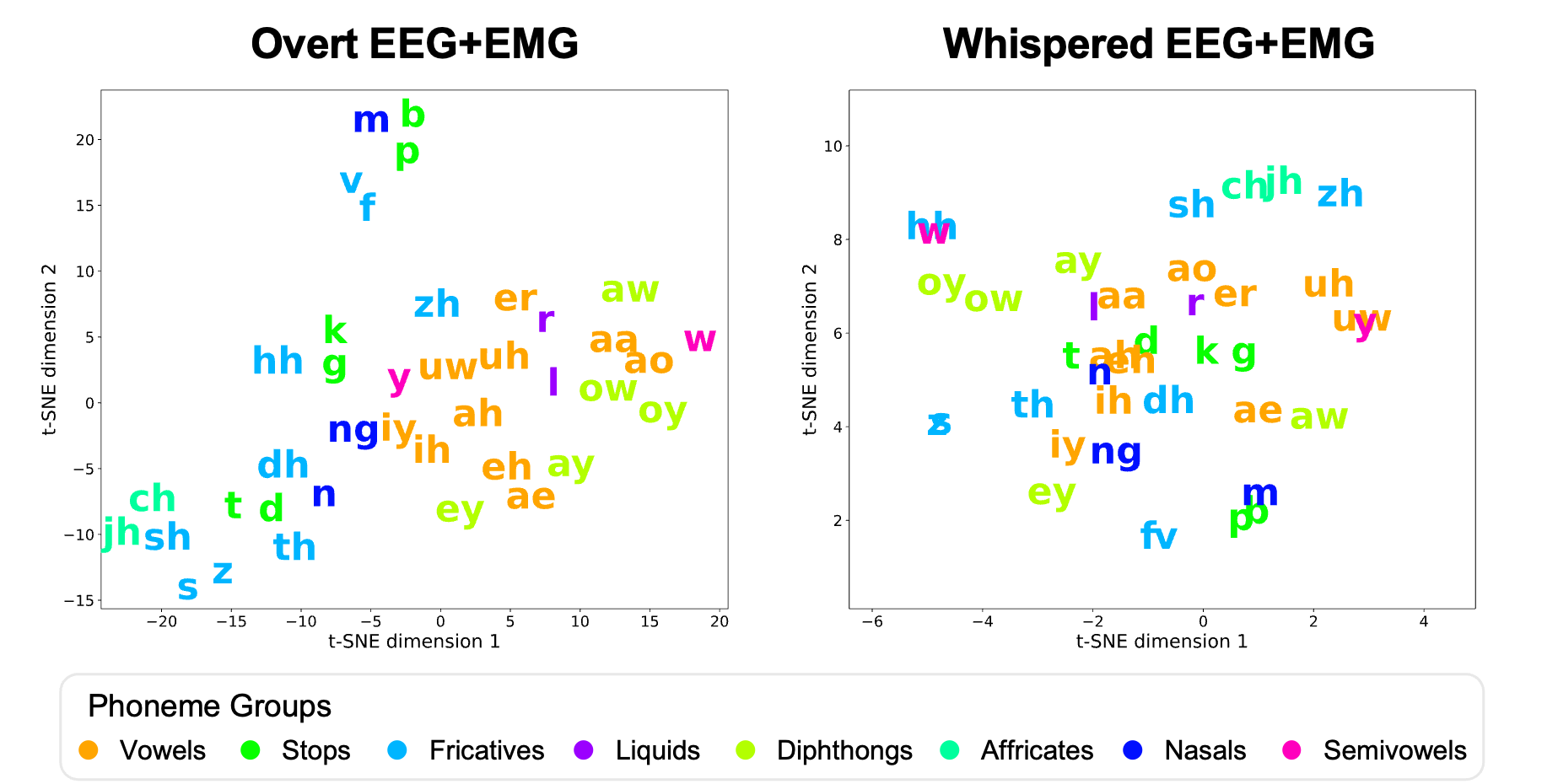}}
\caption{Group-level t-SNE visualization of the phoneme features from model-generated phoneme sequences using EEG combined with EMG signals in overt and whispered modes, across all subjects.}
\label{tsne}
\end{figure}

For the qualitative evaluation, the mel-spectrograms and audio waveforms were presented in Fig.~\ref{mel}, including the original audio alongside the reconstructed voices derived from EEG signals and the combination of EEG and EMG signals for overt and whispered speech. Mel-spectrograms for all 15 subjects are provided in Supplementary Fig.~S2. The reconstructed mel-spectrogram exhibited patterns consistent with those of the original overt speech audio, reflecting key features of silent and vocalized segments. In the case of whispered speech, although the mel-spectrogram of the reconstructed voice exhibits reduced clarity and spectral intensity compared to overt speech, it still displays structural patterns that correspond to the temporal and spectral features of the original overt audio.

Table~\ref{tab2} demonstrates examples of reconstructed sentences, where EEG concatenated with EMG signals was used for both overt and whispered speech. The synthesized audio was processed through a pre-trained ASR model to evaluate the CER and produce text outputs. For overt speech, the reconstructed sentences closely resembled the GT, achieving a low CER of 0.18 and 0.36 for the respective example sample. In whispered speech, while the reconstructed sentences demonstrated slightly higher CER values of 0.5 and 0.4 respectively, the outputs still captured phonological elements of the original sentences. These results highlight the model’s ability to reconstruct intelligible sentences from biosignals, with overt speech providing superior fidelity while whispered speech retains moderate accuracy despite its inherently weaker input signals.

\begin{table}[t!]
\caption{Phoneme Groups Written in ARPAbet}
\centering
\renewcommand{\arraystretch}{1.5}
\begin{tabular}{ll}
\toprule
\textbf{Group} & \textbf{Phonemes of the Group} \\ 
\midrule
Vowels & [iy] [ih] [eh] [ae] [aa] [ao] [uh] [uw] [ah] [er] \\ 
Diphthongs & [ey] [ay] [aw] [oy] [ow] \\ 
Stops & [p] [b] [t] [d] [k] [g] \\ 
Affricates & [ch] [jh] \\ 
Fricatives & [f] [v] [th] [dh] [s] [z] [sh] [zh] [hh] \\ 
Nasals & [m] [n] [ng] \\ 
Liquids & [l] [r] \\ 
Semivowels & [w] [y] \\ 
\bottomrule
\end{tabular}
\label{tab3}
\end{table}

\subsection{Sentence Reconstruction in a Speech-impaired Patient}
To further investigate the applicability of our approach to individuals with speech impairments, we included data from a patient diagnosed with ataxic dysarthria due to cerebellar ataxia (66-year-old male, Korean). The study was approved by the Korea University Institutional Review Board [KUIRB-2023-0429-01]. Due to the patient’s health condition, only 155 sentences were collected through attempted overt speech, with each sentence recorded in a single trial. The dataset was split into 145 training and 10 test sentences, ensuring that the test set consisted of unseen sentences, maintaining consistency with the experimental design.

The results revealed that sentence reconstruction performance in overt speech was slightly lower than that of healthy participants but still demonstrated meaningful decoding results. Specifically, the phoneme accuracy EEG-based decoding achieved 42.85\%, which further increased to 45.63\% when EEG and EMG were combined. The RMSE of EEG-based reconstruction was 0.60, which improved to 0.55 when combining EEG and EMG signals. Also, the MCD values were 4.50 for EEG and 4.10 for EEG combined with EMG, and the F1-score improved from 0.34 with EEG alone to 0.40 with the integration of EEG and EMG signals. These results indicate that multimodal integration improves decoding performance even in patients with speech impairments.

The qualitative results for the patient are illustrated in Supplementary Fig.~S3. The mel-spectrograms of the reconstructed voices from EEG and EEG combined with EMG signals preserved temporal and spectral structures of the original audio, even though the patient exhibited severely impaired articulation. These findings suggest that the proposed framework could be beneficial for patients with ataxic dysarthria and potentially other neuromuscular disorders, offering a viable pathway for adaptive neural communication solutions.

\subsection{Phoneme-level Performance}
Fig.~\ref{confusion} presents the confusion matrix of the predicted phoneme sequence results aggregated across all participants. Fig.~\ref{tsne} shows the group-level t-SNE visualization of the phoneme features based on EEG combined with EMG signals, for all subjects. Supplementary Fig.~S4 further provides the corresponding t-SNE visualization derived from EEG signals alone. In Fig.~\ref{confusion}, particularly when combining EEG and EMG signals, misclassifications predominantly occurred within the same phoneme groups. The red boxes highlight the phoneme groups defined in Table~\ref{tab3}, illustrating this tendency. A similar clustering is also observed in Fig.~\ref{tsne} and Supplementary Fig.~S4, where phoneme features are organized by phoneme groups, demonstrating the model’s ability to capture phoneme-level distinctions.

\begin{table*}[t!]
\caption{Examples of Phoneme Accuracy and Sentence Properties for Whispered EEG Data \\from a Representative Subject with Sentences of Five Best and Five Worst Performance}
\centering
\renewcommand{\arraystretch}{1}
\resizebox{0.9\textwidth}{!}{
\begin{tabular}{cp{10cm}rrr}
\toprule
& Sentence (Phoneme Sequence Length) & Phoneme & W Score & W Score\\ 
& /Phoneme Sequence/ & Accuracy & w/ `sil' & w/o `sil' \\ 
\midrule
\multirow{12}{*}{Best} 
& \textbf{I want to be a princess} (19) &&&\\
& /sil, ay, w, aa, n, t, t, ah, b, iy, ah, p, r, ih, n, s, eh, s, sil/& 63.44 & 0.0496 & 0.0476 \\ 
\cmidrule(lr){2-5} 
& \textbf{Walk on the water} (13) &&&\\
&/sil, w, ao, k, ao, n, dh, ah, w, ao, t, er, sil/ & 52.14 & 0.0445 & 0.0363 \\ 
\cmidrule(lr){2-5} 
& \textbf{I want to be Einstein} (17)&&&\\
& /sil, ay, w, aa, n, t, t, ah, b, iy, ay, n, s, t, ay, n, sil/& 49.45 & 0.0478 & 0.0442 \\ 
\cmidrule(lr){2-5} 
& \textbf{Korea university} (17)&&&\\
&/sil, k, ao, r, iy, ah, y, uw, n, ah, v, er, s, ah, t, iy, sil/& 46.95  & 0.0506 & 0.0478 \\ 
\cmidrule(lr){2-5} 
& \textbf{Which theatre shows ``Mother Goose"?} (23)&&&\\
&/sil, hh, w, ih, ch, th, iy, ah, t, er, sil, sh, ow, z, sil, m, ah, dh, er, g, uw, s, sil/ & 46.41 & 0.0440 & 0.0335 \\ 
\midrule
\multirow{16}{*}{Worst} 
 & \textbf{I want to float in the universe} (24)&&&\\
&/sil, ay, w, aa, n, t, t, ah, f, l, ow, t, ih, n, dh, iy, y, uw, n, ah, v, er, s, sil/& 20.69  & 0.0438 & 0.0423 \\ 
\cmidrule(lr){2-5} 
 & \textbf{Ralph prepared red snapper with fresh lemon sauce for dinner} (47)&&&\\
& /sil, r, ae, l, f, p, r, iy, p, eh, r, d, sil, r, eh, d, s, n, ae, p, er, sil, w, ih, th, sil, f, r, eh, &&&\\
& sh, l, eh, m, ah, n, s, ao, s, sil, f, ao, r, d, ih, n, er, sil/ & 21.16& 0.0432 & 0.0373 \\ 
\cmidrule(lr){2-5} 
& \textbf{Thomas thinks a larger clamp solves the problem} (40)&&&\\
&/sil, t, aa, m, ah, s, sil, th, ih, ng, k, s, sil, ah, l, aa, r, jh, er, k, l, ae, m, p, s, aa, l, v, &&&\\
&z, sil, dh, ah, p, r, aa, b, l, ah, m, sil/ & 21.80 & 0.0453 & 0.0403 \\ 
\cmidrule(lr){2-5} 
& \textbf{Shell shock caused by shrapnel is sometimes cured through group therapy} (56)&&&\\
&/sil, sh, eh, l, sh, aa, k, sil, k, ao, z, d, sil, b, ay, sil, sh, r, ae, p, n, ah, l, sil, ih, z, s, ah,&&&\\
& m, t, ay, m, z, sil, k, y, uh, r, d, sil, th, r, uw, sil, g, r, uw, p, sil, th, eh, r, ah, p, iy, sil/ & 24.04 & 0.0455 & 0.0350 \\ 
\cmidrule(lr){2-5} 
& \textbf{The advertising verse of plymouth variety store never changes} (48)&&&\\ 
&/sil, dh, iy, sil, ae, d, v, er, t, ay, z, ih, ng, v, er, s, sil, ah, v, sil, p, l, ih, m, ah, th, v, &&&\\ 
&er, ay, ah, t, iy, s, t, ao, r, sil, n, eh, v, er, ch, ey, n, jh, ih, z, sil/& 27.02 & 0.0429 & 0.0371 \\ 
\bottomrule
\end{tabular}
}
\label{tab4}
\end{table*}

\begin{figure}[t!]
\centerline{\includegraphics[width=0.945\columnwidth]{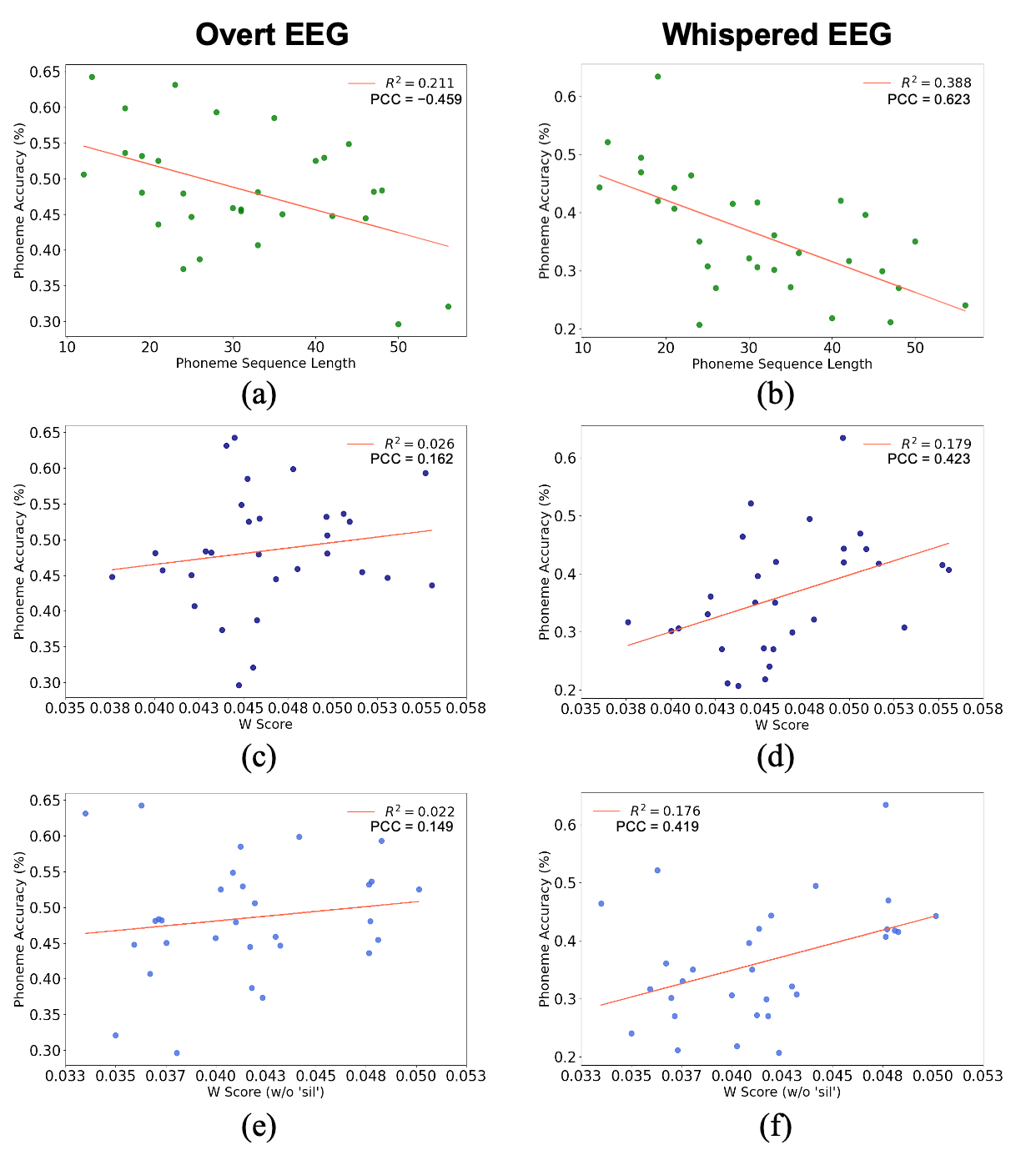}}
\caption{Correlation between phoneme accuracy and phoneme sequence length for overt (a) and whispered EEG data (b), correlation with W scores including `sil' for overt (c) and whispered EEG data (d), and excluding `sil' for overt (e) and whispered EEG data (f), from a representative subject.}
\label{ph}
\end{figure}

\subsection{Properties Affecting Phoneme Decoding Performance in Sentence Reconstruction}
Fig.~\ref{ph} illustrates the relationship between phoneme decoding accuracy and sentence properties for both overt and whispered EEG data. Fig.~\ref{ph}(a) and Fig.~\ref{ph}(b) show a negative correlation between phoneme sequence length and phoneme accuracy for overt EEG with a PCC of $-0.459$, and whispered EEG with $-0.623$, indicating that longer sentences reduce decoding performance, especially for whispered EEG. Fig.~\ref{ph}(c)--(f) show the correlation between phoneme accuracy and W scores with and without `sil'. The results demonstrate that the inclusion or exclusion of `sil' has little impact on the correlation between W scores and decoding performance across speech modes. W scores show a moderate association with decoding performance in whispered EEG and a limited effect in overt EEG, as indicated by the PCC. Specifically, for overt EEG, the PCC is 0.162 when `sil' is included and 0.423 when excluded. For whispered EEG, the PCC is 0.149 with `sil' and 0.419 without `sil'. In addition, Supplementary Fig.~S5 presents the corresponding correlations for overt and whispered EEG combined with EMG data.
Table~\ref{tab4} summarizes the results of phoneme accuracy and sentence properties for whispered speech EEG data, categorizing sentences into the top 5 (best-performance) and bottom 5 (worst-performance) cases. Additional examples for overt EEG, overt EEG combined with EMG, and whispered EEG combined with EMG are provided in Supplementary Tables~S4–S6.

\subsection{Neurophysiological Analysis} 
\begin{figure}[t!]
\centerline{\includegraphics[width=0.88\columnwidth]{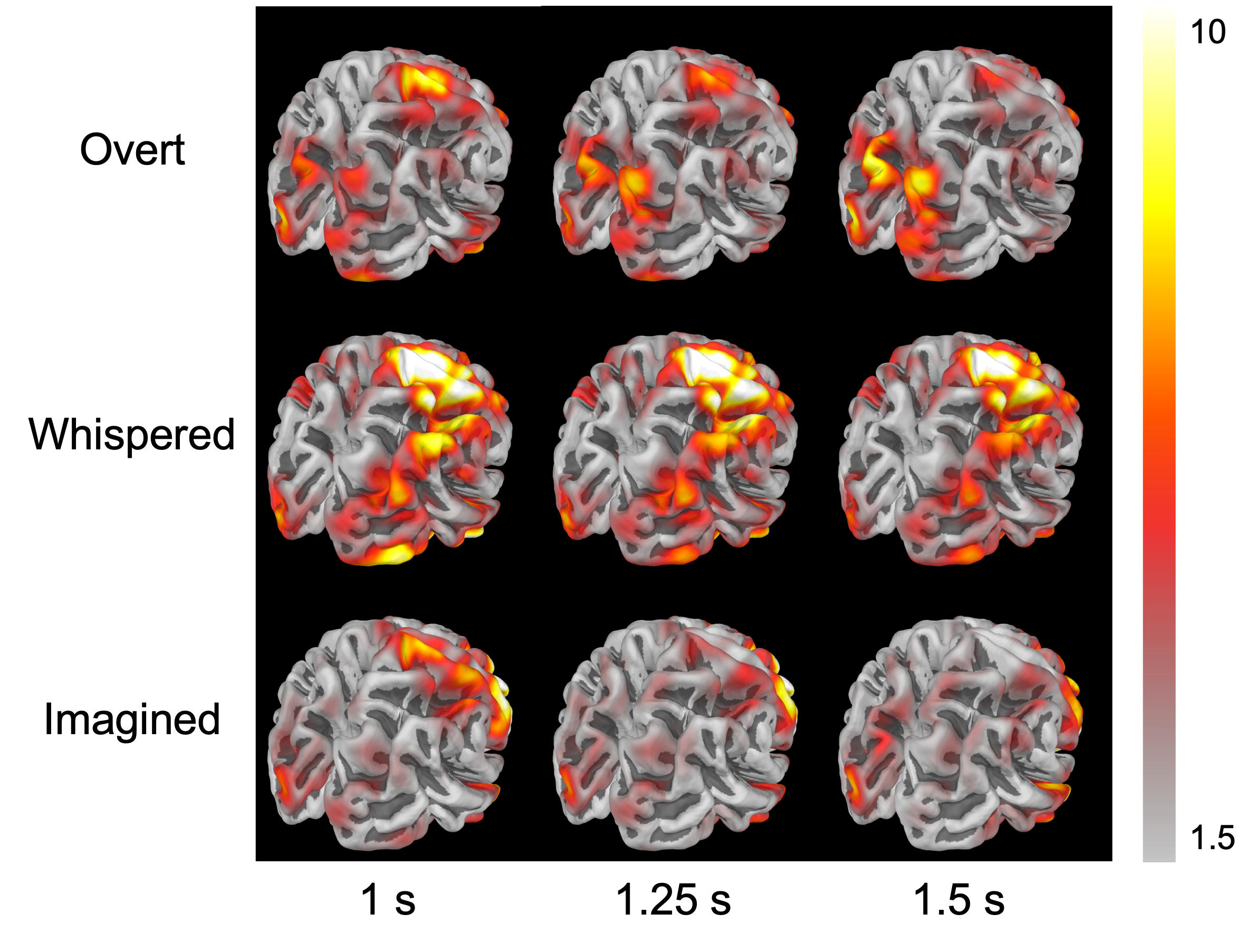}}
\caption{sLORETA visualizations at 1~s, 1.25~s, and 1.5~s after speech onset from a representative subject. Brain activation patterns during overt speech, whispered speech, and imagined speech are displayed in parietal views. The color bar indicates the sLORETA-standardized cortical current density estimates, with higher values reflecting stronger neural activation.}
\label{sl}
\end{figure}

Table~\ref{tab5} details the performance results from unseen sentence decoding across different frequency bands in overt, whispered, and imagined speech EEG. The delta bands exhibited superior performance across all metrics, achieving the highest phoneme decoding accuracy of 36.82\% for overt speech and a competitive accuracy of 31.35\% for imagined speech. In contrast, gamma and high gamma were prominent in imagined speech, showing phoneme accuracies of 31.13\% and 31.35\%, respectively. Fig.~\ref{sl} visualizes the source localization using sLORETA at 1~s, 1.25~s, and 1.5~s after speech onset. Supplementary Fig.~S6 further presents the same sLORETA results in lateral and dorsal views. Activation patterns were observed in the frontal and temporal lobes across all speech modes. Interestingly, imagined speech demonstrated the most pronounced activation in the frontal lobe compared to overt and whispered speech.

\section{Discussion}

\begin{table}[]
\caption{Reconstruction Performance across Frequency Bands \\in Overt, Whispered, and Imagined EEG Data \\from a Representative Subject}
\centering
\renewcommand{\arraystretch}{1.2}
\begin{tabular}{cccccc}
\toprule
\textbf{} & \textbf{} & \textbf{\begin{tabular}[c]{@{}c@{}}Phoneme\\ Accuracy (\%)\end{tabular}} & \textbf{RMSE} & \textbf{MCD}  \\ 
\midrule
\multirow{3}{*}{\textbf{\begin{tabular}[c]{@{}c@{}}Delta\\ (0.5--4 Hz)\end{tabular}}}         
& \textbf{Overt}     & \textbf{36.82} & \textbf{0.53} & \textbf{4.38} \\
& \textbf{Whispered} & 30.98          & \textbf{0.65} & \textbf{5.30} \\
& \textbf{Imagined}  & \textbf{31.35} & \textbf{0.65} & \textbf{5.33} \\ 
\midrule
\multirow{3}{*}{\textbf{\begin{tabular}[c]{@{}c@{}}Theta\\ (4--8 Hz)\end{tabular}}}           
& \textbf{Overt}     & 28.26 & 0.67 & 5.41 \\
& \textbf{Whispered} & 29.60 & 0.75 & 5.98 \\
& \textbf{Imagined}  & 28.61 & 0.71 & 5.99 \\ 
\midrule
\multirow{3}{*}{\textbf{\begin{tabular}[c]{@{}c@{}}Alpha\\ (8--12 Hz)\end{tabular}}}          
& \textbf{Overt}     & 23.67 & 0.69 & 5.67 \\
& \textbf{Whispered} & 30.92 & 0.86 & 6.77 \\
& \textbf{Imagined}  & 31.19 & 0.69 & 5.73 \\ 
\midrule
\multirow{3}{*}{\textbf{\begin{tabular}[c]{@{}c@{}}Beta \\ (12--30 Hz)\end{tabular}}}         
& \textbf{Overt}     & 28.29          & 0.74 & 5.85 \\
& \textbf{Whispered} & \textbf{31.29} & 0.86 & 6.84 \\
& \textbf{Imagined}  & 29.50          & 0.72 & 5.94 \\ 
\midrule
\multirow{3}{*}{\textbf{\begin{tabular}[c]{@{}c@{}}Gamma \\ (30--70 Hz)\end{tabular}}}        
& \textbf{Overt}     & 23.64 & 0.80 & 6.32 \\
& \textbf{Whispered} & 31.21 & 0.76 & 6.07 \\
& \textbf{Imagined}  & 31.13 & 0.69 & 5.69 \\ 
\midrule
\multirow{3}{*}{\textbf{\begin{tabular}[c]{@{}c@{}}High\\ Gamma\\ (70--200 Hz)\end{tabular}}} 
& \textbf{Overt}     & 17.33          & 0.74 & 6.00 \\
& \textbf{Whispered} & 26.23          & 0.74 & 6.02 \\
 & \textbf{Imagined} & \textbf{31.35} & 0.71 & 5.72 \\ 
 \bottomrule
\end{tabular}
\label{tab5}
\end{table}
\subsection{Unseen Sentences Generation for Open Vocabulary Neural Communication}
Reconstructing sentences from speech-related biosignals is significant in providing intuitive means of communication to patients and individuals with diverse needs. The proposed framework demonstrates the feasibility of reconstructing previously unseen sentences using high-density EEG signals, either independently or in combination with EMG data. Unlike word-level decoding, which is often limited in flexibility and offers restricted interaction, sentence-level speech generation could support more natural and comprehensive communication~\cite{wang2022open, amrani2024deep}. Our experimental design comprising various sentences, enabled the training of various phonemes to further reconstruct unseen sentences. The sentences used for training and testing were completely distinct, with the test dataset comprising novel sentences that were not present in the training data. This ensured that the reconstructed outputs were not constrained by predefined sentence categories, providing evidence for the potential of intuitive and flexible open-vocabulary neural communication systems. The confusion matrix and t-SNE visualizations further reveal the model's ability to learn phoneme-level distinctions, with improved true positive rates when combining EEG and EMG signals. The clustering of phonemes within phonetic groups indicates that the model leverages phoneme similarities to enhance prediction accuracy.

\subsection{Significance of Applying Appropriate Modalities and Speech Modes}
As the availability and suitability of specific modalities and speech modes vary depending on the individual's condition, adjusting these choices to individual needs is crucial for developing effective biosignal-based innovative systems for communication and rehabilitation~\cite{mohanchandra2016communication, lazarou2018eeg}. The results highlight several insights for the design and application of neural communication systems in different speech paradigms and clinical scenarios. The integration of EEG and EMG data significantly enhanced sentence reconstruction performance, particularly for overt speech. Metrics such as phoneme accuracy, RMSE, MCD, F1-score showed considerable improvement with EMG integration. This integration offers a promising avenue for improving communication outcomes in individuals with partial speech and muscle function. Even in whispered speech, where the input signals are inherently weaker, phoneme accuracy and F1-score improved with the addition of EMG data, highlighting the utility of EMG as a supplementary modality when muscle activity is available. Additionally, the mel-spectrograms of reconstructed whispered speech closely resembled those of overt speech, underscoring its applicability in conditions where overt speech is not feasible, such as in cases of vocal cord impairment or for patients capable only of whispered speech. Moreover, the framework demonstrated the potential for reconstructing previously unseen sentences using EEG alone, offering a viable pathway for non-invasive neural communication in scenarios where EMG signals are inaccessible, such as in cases of severe paralysis.

\subsection{Influence of Sentence Properties on Reconstruction Performance}
Interestingly, sentence length emerged as a critical factor influencing the performance in sentence reconstruction tasks, with shorter sentences generally outperforming longer ones. This observation diverges from previous findings in classification tasks, where longer sequences have been shown to enhance performance due to the higher complexity of the words~\cite{nguyen2017inferring}. The impact of sentence length on performance was more pronounced in whispered speech compared to overt speech. This may be attributed to the differences in signal quality and information density between the two speech modes. Overt speech brain signals may carry richer phonological and acoustic features produced during actual vocalization and attempts to move the articulators, enabling more effective decoding, even for longer sentences.

These findings provide insight into the need for optimized approaches for each speech mode, particularly for whispered speech, to enhance decoding performance, especially for longer sentences.

\subsection{Neurophysiological Insights Across Speech Paradigms}
The findings from this study provide valuable insights into the frequency-specific and hierarchical neural dynamics underlying speech decoding across overt, whispered, and imagined speech paradigms. High-frequency bands exhibited prominence in imagined speech, reflecting their association with auditory imagery and inner speech representation~\cite{lu2021neural}. Also, delta band activity provided a temporal framework for sentence-level linguistic processing, aligning cortical dynamics with the rhythmic constructs essential for speech generation~\cite{lu2021neural}. These results are in line with prior research, suggesting that delta oscillations establish a temporal scaffold, while gamma activity facilitates phonetic and semantic encoding, together enabling imagined speech decoding~\cite{proix2022imagined}. Also, the increased activation in the frontal lobe during imagined speech reflects its reliance on higher-order cognitive processes, differentiating it from overt and whispered speech modes~\cite{zhang2024chisco}.  The consistent involvement of the temporal lobe across all modes emphasizes its critical role in integrating auditory and linguistic information, providing a stable foundation for decoding speech signals across diverse contexts~\cite{crinion2003temporal}.

\subsection{Potential for Silent Communication}

Imagined speech demonstrated the most pronounced activation in the frontal lobe, suggesting a greater reliance on cognitive and planning-related processes compared to overt and whispered speech. This aligns with the findings in the previous studies that imagined speech engages abstract and higher-order linguistic representations without explicit motor articulation~\cite{zhang2024chisco}. Imagined speech holds unique promise for silent communication technologies, particularly for individuals with complete speech or motor function loss, as it bypasses the need for articulation. To optimize decoding performance, it is essential to consider the distinct neural activation patterns of different speech modes and strategically utilize frequency-specific neural activity. Despite promising results in this study for generating unseen, open-vocabulary, and sentence-level speech from non-invasive biosignals, challenges remain, particularly in accurately reconstructing imagined speech. Future research should explore adaptive models, such as self-supervised learning, to address signal variability and enhance robustness. Expanding datasets to include more diverse and longer sentences and improving signal quality and decoding frameworks will be critical for advancing the reconstruction of unconstrained sentences in imagined speech. In addition, since the speech modes in our experimental paradigm were presented in a fixed order from overt to whispered to imagined, it is possible that echoic trace effects from earlier modes may have subtly influenced performance in the imagined speech condition, and this issue may be explored in future research.

\section{Conclusion}

In this study, we demonstrated the feasibility of reconstructing unconstrained speech by leveraging phoneme-level information extracted from human biosignals. The results demonstrate that integrating speech-related biosignals for each mode of speech significantly enhances decoding performance and speech reconstruction quality. Moreover, the possibility of reconstructing speech using EEG signals alone was suggested. These findings emphasize the potential of open-vocabulary neural communication systems based on non-invasive biosignals, accommodating the specific needs of individuals with varying degrees of speech impairment and conditions.
Furthermore, several insights gained from this study contribute to the further development of EEG-based decoding systems to address a wide range of communication impairments. Future research should prioritize overcoming the challenges posed by low signal quality in whispered and imagined speech, particularly for fully paralyzed individuals. Our study may provide groundwork for intuitive, non-invasive, and adaptive BTS systems that have the potential to transform communication for individuals with severe speech limitations. In addition, reconstructed audio samples and a toy dataset from a single subject are available at https://deokseonkim.github.io/eeg2speech/.

\section*{Acknowledgment}

The authors sincerely thank Jun-Young Kim for his insightful discussions, which significantly contributed to development and completion of this research.

\end{document}